\documentclass[twocolumn,showpacs,preprintnumbers,amsmath,amssymb]{revtex4}
\usepackage{graphicx,subfigure,textcomp}% Include figure files
\usepackage{dcolumn}% Align table columns on decimal point
\usepackage{bm}% bold math
\usepackage{mathrsfs,slashed,amsmath}
\usepackage{natbib}
\allowdisplaybreaks
\begin{document}
\title{The first PDF moments for three dynamical flavors in BChPT}
\author{Peter C. Bruns}
\affiliation{Institut f\"ur Theoretische Physik, Universit\"at Regensburg, D-93040 Regensburg, Germany}
\author{Ludwig Greil}
\affiliation{Institut f\"ur Theoretische Physik, Universit\"at Regensburg, D-93040 Regensburg, Germany}
\author{Andreas Sch\"afer}
\affiliation{Institut f\"ur Theoretische Physik, Universit\"at Regensburg, D-93040 Regensburg, Germany}
\date{\today}
\begin{abstract}
We present a calculation of generalized baryon form factors in the framework of three-flavor covariant baryon chiral perturbation theory at leading one-loop order, necessary for the calculation of the first moments of generalized parton distribution functions. The formulae we derive can be used to guide the chiral extrapolation of three-flavor lattice calculations of the corresponding QCD matrix elements.
\end{abstract}
\maketitle
\section{Introduction}
The concept of generalized parton distributions (GPDs) connects several different, seemingly unconnected hadron structure observables such as form factors, angular momentum carried by quarks and gluons, moments of parton distribution funtions, transverse spatial structure, etc.. Thus, GPDs and Mellin moments thereof are important quantities for studying hadron structure~\cite{Ji:1996ek,Diehl:2003ny}.\\
Lattice QCD has proven to be able to probe hadron structure in the non-perturbative regime quite successfully for two dynamical quark flavors \cite{Hagler:2009ni}, but today most simulations are carried out for three or four   dynamical flavors. The quark masses in these simulations are typically unphysically large as it often seems more efficient to invest computing time in simulating with larger volumes, finer lattices or better fermion actions. Covariant baryon chiral perturbation theory (BChPT) \cite{Gasser:1987rb,Bernard:2007zu} is an effective field theory of QCD which supplies extrapolation formulae for variable quark masses (pseudoscalar meson masses) which are of vital importance to thoroughly analyze lattice data.\\
In this article, we generalize the work presented in \cite{Dorati:2007bk} to three quark flavors. In sect. 2 we shortly review the connection between GPDs and the so-called generalized form factors. In sect. 3 we present the effective chiral Lagrangian needed for a leading one-loop calculation of the $SU(3)$ version of $A^{s,v}_{2,0}$ in the forward limit and in sect. 4 we show the results for the nucleon sector. We give a short conclusion in sect. 5.
\section{GPDs and generalized form factors}
It is well known that the parity-even generalized parton distribution functions $H^q(x,\xi,\Delta^2)$ and $E^q(x,\xi,\Delta^2)$ (with Bjorken-$x$, skewedness $\xi$ and momentum transfer $\Delta$) are connected to three generalized form factors $A^q_{2,0}(\Delta^2)$, $B^q_{2,0}(\Delta^2)$ and $C^q_{2,0}(\Delta^2)$ via their first Mellin moments \cite{Diehl:2003ny}. These generalized form factors themselves are accessible through calculation of baryon matrix elements of totally symmetrized and traceless local operators 
\begin{align}
\mathcal{O}^q_{\mu\nu}=i\bar{q}\gamma_{\{\mu}\overleftrightarrow{D}_{\nu\}}q\,.
\end{align}
Here, we have introduced the abbreviations
\begin{align}
A_{\{\mu}B_{\nu\}}&=\frac{1}{2}\left(g_{\alpha\mu}g_{\beta\nu}+g_{\beta\mu}g_{\alpha\nu}-\frac{2}{d}g_{\alpha\beta}g_{\mu\nu}\right)A^{\alpha}B^{\beta},\\
\overleftrightarrow{D}_{\mu}&=\frac{1}{2}\left(\overrightarrow{D}_{\mu}-\overleftarrow{D}_{\mu}\right),\end{align}
where $d$ represents the space-time dimension. For two light quark flavors and assuming isospin symmetry, this matrix element can be decomposed into said generalized form factors. In the $SU(3)$ case, however, one finds five different form factors if one imposes no restrictions on isospin or baryon content. In this work, we analyze the flavor-singlet and the flavor-octet sector, \textit{i.e.}
\begin{align}
\mathcal{M}_{B'B}^s&=\langle B',s',\mathbf{p}'|i\bar{q}\openone\gamma_{\{\mu}\overleftrightarrow{D}_{\nu\}}q| B,s,\mathbf{p}\rangle,\label{eq:Ms}\\ 
\mathcal{M}_{B'B}^{v,i}&=\langle B',s',\mathbf{p}'|i\bar{q}\lambda^i\gamma_{\{\mu}\overleftrightarrow{D}_{\nu\}}q| B,s,\mathbf{p}\rangle, \label{eq:Mv}
\end{align}
where the $\lambda^i$ denote the Gell-Mann matrices and $B$($B'$) labels the incoming (outgoing) baryon from the lowest-lying baryon octet. The decomposition of these matrix elements yields the following result:
\begin{align}
\begin{split}\label{eq:decomp}
\mathcal{M}_{B'B}^{s,v}&=\bar{u}(p')\biggl[A^{s,v}_{B'B}(\Delta^2)\gamma_{\{\mu}\bar{p}_{\nu\}}\\
&\quad-i\frac{B^{s,v}_{B'B}(\Delta^2)}{2\bar{m}}\Delta^{\alpha}\sigma_{\alpha \{\mu}\bar{p}_{\nu\}}+\frac{C^{s,v}_{B'B}(\Delta^2)}{\bar{m}}\Delta_{\{\mu}\Delta_{\nu\}}\\
&\quad+\frac{D^{s,v}_{B'B}(\Delta^2)}{2\bar{m}}\bar{p}_{\{\mu}\Delta_{\nu\}}+E^{s,v}_{B'B}(\Delta^2)\gamma_{\{\mu}\Delta_{\nu\}}\biggr]u(p).
\end{split}
\end{align}
Here, we have introduced another standard momentum variable $\bar{p}=(p'+p)/2$. Moreover, $\bar{m} = (m_{B}+m_{B'})/2\,$. Taking $\mathcal{P}$-, $\mathcal{C}$- and $\mathcal{T}$-symmetry into account, we find that $A^{s,v}$, $B^{s,v}$ and $C^{s,v}$ are Hermitian $8\times8$-matrices, whereas $D^{s,v}$ and $E^{s,v}$ are anti-Hermitian. These matrices are directly accessible via $SU(3)$ BChPT. \\
As also mentioned in \cite{Diehl:2003ny}, in the forward limit, the generalized form factor $A^q_{2,0}(0)$ is linked to the first moment of the parton distribution functions (PDFs) $q(x)$ and $\bar{q}(x)$ via the relation
\begin{align}
\langle x \rangle^q =\int_0^1dx\,x\left[q(x)+\bar{q}(x)\right],
\end{align}
and thus, certain $A^i_{B'B}(0)$ or linear combinations thereof are connected to a linear combination of these $\langle x \rangle^q$, \textit{e.g.} for the isovector moment, we find
\begin{align}
A^{v,3}_{pp}(0)=\frac{1}{2}\left(\langle x \rangle^u-\langle x \rangle^d\right)\equiv\frac{1}{2}\langle x\rangle_{u-d}\,.
\end{align}
\section{Effective Lagrangians}
Chiral Perturbation Theory provides low-energy expansions of QCD Green functions in terms of a small parameter $p$ (small compared to a typical hadronic scale of $\sim$1 GeV), where $p$ can stand for meson four-momenta, baryon three-momenta or meson masses. 
The well-known leading order $SU(3)$ Lagrangian \cite{Krause:1990xc} in the one-baryon-sector reads
\begin{align}
\begin{split}
\mathscr{L}^{(1)}_{MB}&=i\langle\bar{B}\gamma^{\mu}[D_{\mu},B]\rangle-m_0\langle\bar{B}B\rangle\\
                         &\quad+\frac{D}{2}\langle\bar{B}\gamma^{\mu}\gamma_5\{u_{\mu},B\}\rangle+\frac{F}{2}\langle\bar{B}\gamma^{\mu}\gamma_5[u_{\mu},B]\rangle,
\end{split}
\end{align}
where $B$ denotes the baryon octet
\begin{align}
B=\begin{pmatrix}\frac{1}{\sqrt{2}}\Sigma^0+\frac{1}{\sqrt{6}}\Lambda & \Sigma^+ & p\\\Sigma^- & -\frac{1}{\sqrt{2}}\Sigma^0+\frac{1}{\sqrt{6}}\Lambda & n \\ \Xi^- & \Xi^0 & -\frac{2}{\sqrt{6}}\Lambda\end{pmatrix},
\end{align}
$m_0$ represents the baryon mass in the chiral limit and $D$ and $F$ are the two axial-vector coupling constants. Furthermore, we collect the pseudoscalar fields in a $3\times3$ unitary matrix $u$ which is defined as $u=\exp\{i\Phi/F_0\}$\,, where
\begin{align}
\Phi=\frac{1}{\sqrt{2}}\begin{pmatrix}\frac{1}{\sqrt{2}}\pi^0+\frac{1}{\sqrt{6}}\eta & \pi^+ & K^+\\\pi^- & -\frac{1}{\sqrt{2}}\pi^0+\frac{1}{\sqrt{6}}\eta & K^0 \\ K^- & \bar{K}^0 & -\frac{2}{\sqrt{6}}\eta\end{pmatrix},
\end{align}
contains the pseudoscalar octet of (pseudo-) Goldstone bosons associated with the spontaneously broken approximate chiral symmetry of QCD, and $F_0$ is the pertinent meson decay constant in the chiral limit. $D_{\mu}$ is the appropriate covariant derivative and is defined as $D_{\mu}B=\partial_{\mu}B+[\Gamma_{\mu},B]\,$, with the chiral connection (we set external vector or axial-vector fields to zero here)
\begin{align}
\Gamma_{\mu}=\frac{1}{2}[u^{\dagger},\partial_{\mu}u]\,.
\end{align}
The operator $u_{\mu}$ is defined as $u_{\mu}=iu^{\dagger}\partial_{\mu}u-iu\partial_{\mu}u^{\dagger}$, again without external vector or axial-vector fields, and transforms in the same way under chiral rotations as the matter-field $B$  \cite{Krause:1990xc,Scherer:2002tk}.\\
As proposed in \cite{Dorati:2007bk}, we now extend this Lagrangian to the interaction between the baryon octet and symmetric, traceless external flavor-singlet $\tilde{v}_{\{\mu\nu\}}$ and flavor-octet tensor fields $v^i_{\{\mu\nu\}}$ of definite parity. For the construction of the Lagrangian, we utilize the tensor structures
\begin{align}
V_{\mu\nu}^{\pm}&=\left(u^{\dagger}\frac{\lambda^i}{2} v^i_{\{\mu\nu\}}u\pm u\frac{\lambda^i}{2} v^i_{\{\mu\nu\}}u^{\dagger}\right)\,,\\
V^0_{\mu\nu}&=\tilde{v}_{\{\mu\nu\}}\times\frac{\openone}{2}\,.
\end{align}
We do not assign a chiral power to these tensor structures and thus our leading order Lagrangian is of zeroth order. We can now write down all terms that are allowed by the symmetry properties of QCD and Lorentz invariance, which leads to the following result:
\begin{align}
\begin{split}
\mathscr{L}_{MB,t}^{(0)}&=\frac{a_D}{4}\langle\bar{B}i\gamma^{\mu}\{V^+_{\mu\nu},D^{\nu}B\}\rangle +\text{h.c.}\\
                        &\quad+\frac{a_F}{4}\langle\bar{B}i\gamma^{\mu}[V^+_{\mu\nu},D^{\nu}B]\rangle +\text{h.c.}\\
                        &\quad+\frac{\Delta a_D}{4}\langle\bar{B}i\gamma^{\mu}\gamma_5\{V^-_{\mu\nu},D^{\nu}B\}\rangle +\text{h.c.}\\
                        &\quad+\frac{\Delta a_F}{4}\langle\bar{B}i\gamma^{\mu}\gamma_5[V^-_{\mu\nu},D^{\nu}B]\rangle +\text{h.c.}\\
                        &\quad+\frac{a_s}{2}\langle\bar{B}i\gamma^{\mu}V^0_{\mu\nu}D^{\nu}B\rangle +\text{h.c.}
\end{split}
\end{align}
For the construction of the $\mathcal{O}(p^2)$ Lagrangian we need additional building blocks, namely the chiral symmetry breaking term $\chi$ which contains the quark mass matrix $\mathcal{M}$:
\begin{align}
\chi&=2B_0\mathcal{M},&\quad \mathcal{M}&=\text{diag}\left(m_u,m_d,m_s\right),\\
\chi_+&=u^{\dagger}\chi u^{\dagger}+u\chi^{\dagger}u,&\quad \tilde{\chi}_+&=\chi_+-1/3\langle\chi_+\rangle.
\end{align}
Using these definitions we can construct the second order Lagrangian needed for our leading-loop calculation:
\begin{align}
\begin{split}\label{eq:Lag2}
\mathscr{L}_{MB,t}^{(2)}&=t_1\langle\bar{B}i\gamma^{\mu}\{V^+_{\mu\nu},D^{\nu}B\}\rangle\langle\chi_{+}\rangle+\text{h.c.}\\
                        &\quad+t_2\langle\bar{B}i\gamma^{\mu}[V^+_{\mu\nu},D^{\nu}B]\rangle\langle\chi_{+}\rangle+\text{h.c.}\\
                        &\quad+t_3\langle\bar{B}i\gamma^{\mu}\{\{V^+_{\mu\nu},\tilde\chi_{+}\},D^{\nu}B\}\rangle+\text{h.c.}\\
                        &\quad+t_4\langle\bar{B}i\gamma^{\mu}[\{V^+_{\mu\nu},\tilde\chi_{+}\},D^{\nu}B]\rangle+\text{h.c.}\\
                        &\quad+t_5\langle\bar{B}i\gamma^{\mu}\{[V^+_{\mu\nu},\tilde\chi_{+}],D^{\nu}B\}\rangle+\text{h.c.}\\
                        &\quad+t_6\langle\bar{B}i\gamma^{\mu}[[V^+_{\mu\nu},\tilde\chi_{+}],D^{\nu}B]\rangle+\text{h.c.}\\
                        &\quad+t_7\langle\bar{B}i\gamma^{\mu}\{V^+_{\mu\nu},\{\tilde\chi_{+},D^{\nu}B\}\}\rangle+\text{h.c.}\\
                        &\quad+t_8\langle\bar{B}i\gamma^{\mu}[V^+_{\mu\nu},\{\tilde\chi_{+},D^{\nu}B\}]\rangle+\text{h.c.}\\
                        &\quad+t_9\langle\bar{B}i\gamma^{\mu}D^{\nu}B\rangle\langle V^+_{\mu\nu}\tilde\chi_{+}\rangle+\text{h.c.}\\
                        &\quad+t_{10}\langle\bar{B}i\gamma^{\mu}V^+_{\mu\nu}\rangle\langle\tilde\chi_{+}D^{\nu}B\rangle+\text{h.c.}\\
                        &\quad+t_{11}\langle\bar{B}i\gamma^{\mu}\{V^0_{\mu\nu},D^{\nu}B\}\rangle\langle\chi_{+}\rangle+\text{h.c.}\\
                        &\quad+t_{12}\langle\bar{B}i\gamma^{\mu}\{\{V^0_{\mu\nu},\tilde\chi_{+}\},D^{\nu}B\}\rangle+\text{h.c.}\\
                        &\quad+t_{13}\langle\bar{B}i\gamma^{\mu}[\{V^0_{\mu\nu},\tilde\chi_{+}\},D^{\nu}B]\rangle+\text{h.c.}\\
\end{split}
\end{align}
Note that the coupling $t_{10}$ multiplies a non-hermitian strucure and can therefore be complex in general. Also, this coupling will give a contribution to the form factor $E$ defined in eq.~(\ref{eq:decomp}).
\section{Results}\label{sec:res}
\begin{figure}[ht]
\centering
\subfigure[]
{\includegraphics[width=0.15\textwidth]{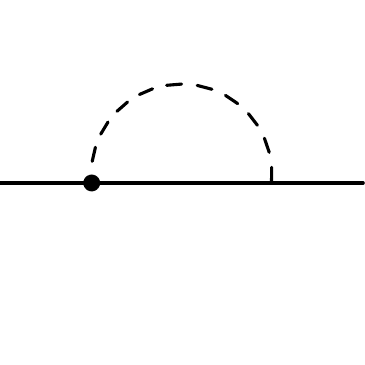}\label{fig:ga}}\quad\subfigure[]{\includegraphics[width=0.15\textwidth]{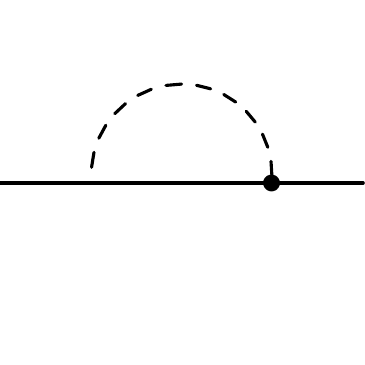}\label{fig:gb}}\quad\subfigure[]{\includegraphics[width=0.15\textwidth]{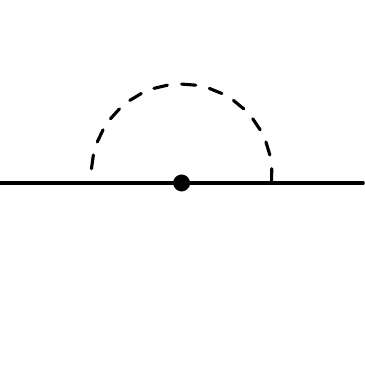}}\quad\subfigure[]{\includegraphics[width=0.15\textwidth]{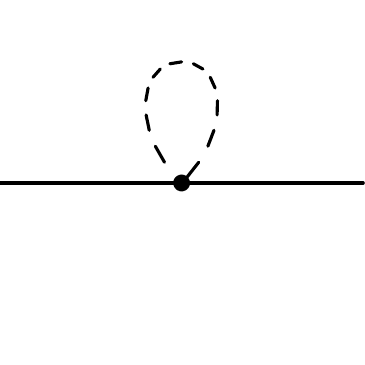}}
\caption{Feynman diagrams contributing to leading-loop order: the solid line denotes baryons, the dashed line denotes pseudoscalar mesons and the dot represents an external tensor field from the $\mathcal{O}(p^0)$ Lagrangian coupling to the baryons and mesons. The tree-level diagrams are not shown here.}
\label{fig:diagrams}
\end{figure}
\begin{figure}
\includegraphics[width=0.45\textwidth]{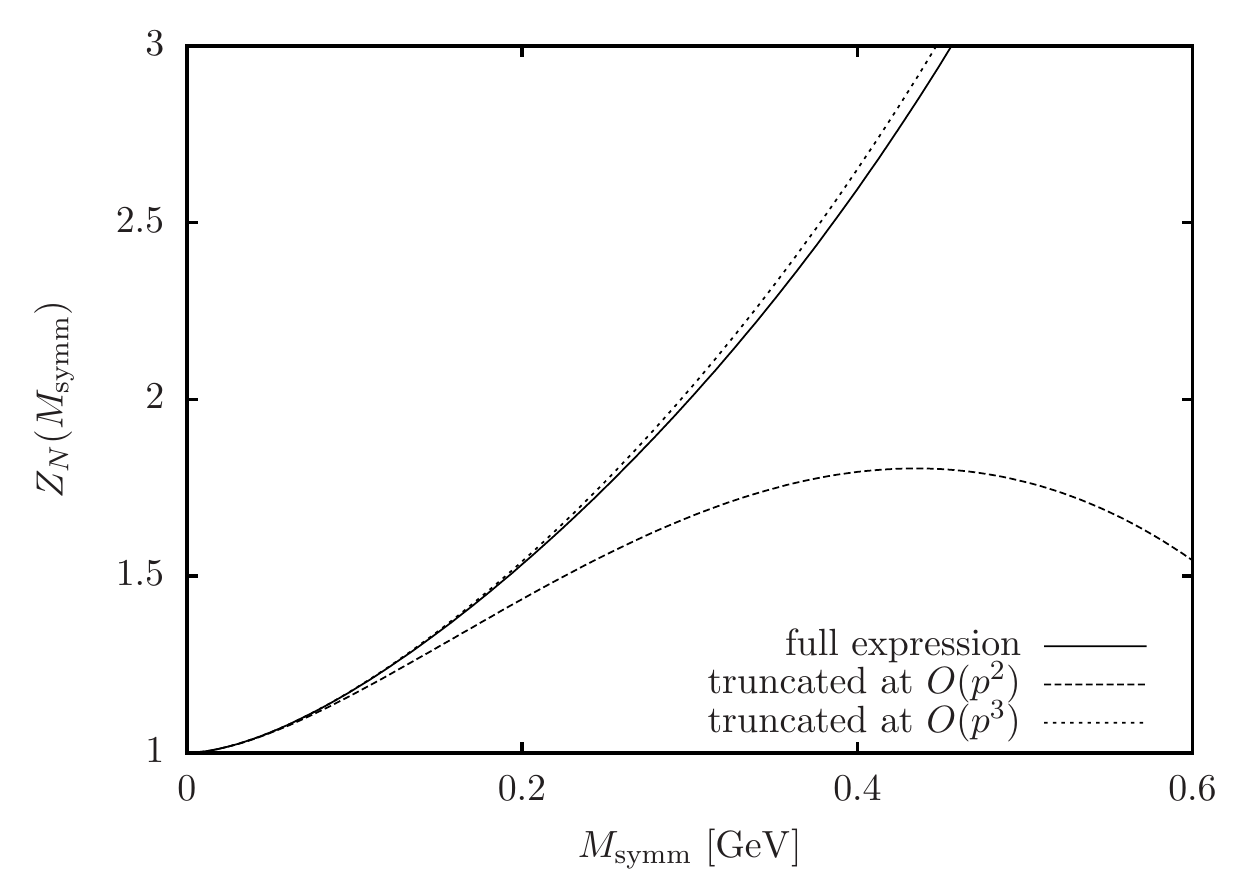}
\caption{One-loop wave function renormalization factor for $M_{\pi}=M_{K}=M_{\eta}\equiv M_{symm}$: full result (solid line) and truncated results (dashed and dotted lines).}
\label{fig:zfac}
\end{figure}

As was mentioned in the introduction, our calculation is intended as a direct generalization of \cite{Dorati:2007bk} to the three-flavor case. Therefore, the only fields appearing as dynamical degrees of freedom are the mesons and baryons from the ground-state octets of flavor-$SU(3)$. The effects of all other fields are (at least to the order we are working here) encoded in the (so far undetermined) LECs appearing in the effective Lagrangians constructed above, in analogy to the treatment of the baryon masses in \cite{Borasoy:1996bx, Frink:2004ic}. Of course, the problem is more urgent in a three-flavor calculation: The threshold energy for e.g. a $K\Sigma$ state, which is included in our calculation, is at about 1685 MeV, while that of $\pi\Delta$ is at 1370 MeV, the $\pi$-Roper threshold at roughly 1580 MeV, and so on. Let us concentrate on the decuplet, which would certainly be the most important resonant state due to the small $N-\Delta$ mass splitting. It has been incorporated in many studies employing three-flavor HBChPT, see e.g. \cite{Jenkins:1990jv, Jenkins:1991es}, and an extension of Infrared Regularization to the case of dynamical decuplet fields has also been given in the meantime \cite{Bernard:2003xf}. For the observables calculated here, however, the inclusion of explicit decuplet fields would not render the theory more effective, since again a host of new undetermined parameters would have to be introduced (e.g. $V\Delta MB$, $V\Delta\Delta$ couplings, where $V$ denotes the tensor source field), of which there are already enough to absorb the leading effects of resonances in our present framework. It might therefore happen that some of our LECs will be enhanced (comparing to natural size estimates) due to important resonance contributions. On the other hand, in order to study the momentum dependence of the form factors, and in particular for the study of finite volume effects, the explicit inclusion of the decuplet fields will probably be inevitable in a three-flavor calculation. These considerations are still work in progress.\\
The standard BChPT power counting formula \cite{Bernard:2007zu} can be written down as
\begin{align}
D=2L+1+\sum_{n}(n-2)N_M^{(n)}+\sum_{n}(n-1)N_{MB}^{(n)}\,,
\end{align}
where $D$ represents the chiral dimension of a Feynman diagram, $L$ denotes the number of loops and $N^{(n)}_{M,MB}$ counts the number of vertices that stem from the meson and meson-baryon Lagrangians of power $p^n$ respectively. In our case, one has to take into account the fact that the lowest order meson-baryon Lagrangian including symmetric tensor fields starts at order $\mathcal{O}(p^0)$ and thus, all leading one-loop order diagrams are of $\mathcal{O}(p^2)$. These topologies are depicted in fig. \ref{fig:diagrams}.\\
From our calculation, we can extract the flavor-singlet and flavor-octet generalized form factor $A^{s,v}_{B'B}$ to $\mathcal{O}(p^2)$ in the forward case for different $B'$ and $B$. In this work, we only show the results for the nucleon-channels, because all expressions are rather lengthy and the complete list is only of interest for practitioners of such calculations. The complete set of results is, however, available electronically \cite{Bruns:2011sh}.  We note that some of these matrix elements have also been calculated in the framework of partially quenched ChPT, see refs. \cite{Chen:2001yi, Detmold:2005pt}.
Our results are exact to leading one-loop order $\mathcal{O}(p^2)$\,. In our formulae, however, we display the full loop functions, which also contain terms of higher order in the meson mass (quark mass) expansion. We have observed that the truncated leading-one-loop results approximate the full loop functions rather badly for meson masses already above $\sim 300\,$ MeV. In particular, the nonanalytic $M^{3}$-contributions to the loops are far from negligible. As an example, we show the baryon one-loop wave function renormalization factor at $M_{\pi}=M_{K}=M_{\eta}\equiv M_{symm}$ in fig. \ref{fig:zfac}, and the truncated $\mathcal{O}(p^2)$ (dashed) and $\mathcal{O}(p^3)$ (dotted) approximations to it. For the other graphs, the situation is very similar. Therefore, to take the numerically important higher-order parts into account, we insist to use the full loop functions everywhere, in accord with the original proposal in \cite{Becher:1999he}. As a side remark, we note that the fact that $Z\gg 1$ for meson masses much above $400\,$MeV may cast doubts on the applicability of BChPT in this regime. Projecting our effective Lagrangian onto the $SU(2)$ sector, and comparing our results for the one-loop amplitudes  with those derived in \cite{Dorati:2007bk}, we also obtain matching relations for the $SU(2)$ parameters $a_{2,0}^{v,s}$ and $c_{8,9}^{(r)}$ to the $SU(3)$ parameters $a_D$, $a_F$, $a_{s}$ and $t_{i}^{(r)}$\,.\\
The results for the generalized form factors $A^{s,v}_{B'B}(0)$ are presented in eqs. \eqref{eq:first}-\eqref{eq:last}, while the matching relations can be found in eqs. \eqref{eq:matfirst}-\eqref{eq:matlast}. We have left out the index $r$ of the renormalized couplings $t_{i}^{(r)}$ for better legibility. In the matching relations, one could also replace the strange quark mass by the Gell-Mann-Oakes-Renner-relation $2B_{0}m_{s}=2M_{K}^{2}-M_{\pi}^{2}$ to the order we are working in. Note that the singlet form factors do not get contributions from loop graphs at leading one-loop order. Also, note that $A^{v,1}_{np} = A^{v,3}_{pp}\,, A^{v,2}_{np} =iA^{v,3}_{pp}\,, A^{v,3}_{nn} = -A^{v,3}_{pp}\,, A^{v,8}_{nn} =A^{v,8}_{pp}$.
\begin{widetext}
\begin{align}
\begin{split}
a^{v}_{2,0} &= a_{D}+a_{F} + 16B_{0}m_{s}(t_{1}+t_{2}-\frac{1}{3}(2(t_{3}+t_{4})-t_{7}-t_{8})) \\ 
          &\quad+\frac{M_{K}^2}{48\pi^{2}F_{0}^{2}} \biggl[a_{F}(D^{2}-6DF-3F^{2})+a_{D} \left(-7 D^2+2 D F-3 F^2\right)\\ 
          &\quad- 3 \left(a_{D} \left(1+7 D^2-2 D F+3 F^2\right)+a_{F} \left(1-D^2+6 D F+3 F^2\right)\right)\log\left(\frac{M_{K}}{m_{0}}\right)\biggr]+\mathcal{O}(M_{K}^{3}),\label{eq:matfirst}
\end{split}\\
\begin{split}
a^{s}_{2,0} &= a_{s}+\frac{16}{3}B_{0}m_{s}(3t_{11}+t_{12}-3t_{13}) +\mathcal{O}(M_{K}^{3}),
\end{split}\\
\begin{split}
c_{8}^{r}(\lambda=m_{0}) &= m_{0}^{2}\left(4(t_{1}+t_{2})+\frac{2}{3}(2(t_{3}+t_{4})-t_{7}-t_{8})\right) + \mathcal{O}(M_{K}^{2}),
\end{split}\\
\begin{split}
c_{9}&= \frac{2}{3}m_{0}^{2}(6t_{11}-t_{12}+3t_{13}) +\mathcal{O}(M_{K}^{2}).\label{eq:matlast}
\end{split}\\
\begin{split}\label{eq:first}
A^{s}_{pp}(0) &= \frac{a_{s}}{2}+\frac{8}{3} M_{K}^2 (3 t_{11}+t_{12}-3 t_{13})+M_{\pi}^2 \left(4 t_{11}-\frac{8 t_{12}}{3}+8t_{13}\right),
\end{split}\\
\begin{split}
A^{s}_{nn}(0) &= \frac{a_{s}}{2}+\frac{8}{3} M_{K}^2 (3 t_{11}+t_{12}-3 t_{13})+M_{\pi}^2 \left(4 t_{11}-\frac{8 t_{12}}{3}+8t_{13}\right),
\end{split}\\
\begin{split}
A^{v,3}_{pp} &= Z_{N}\frac{a_{D}+a_{F}}{2}+ \frac{4}{3} \biggl[ 3 (t_{1}+t_{2}) (2 M_{K}^2 + M_{\pi}^2)-2 (2 t_{3} + 2 t_{4} - t_{7} - t_{8}) (M_{K}^2 - M_{\pi}^2)\biggr] \\ 
&\quad+ \frac{I_{M}(M_{\pi})}{(24 F_{0}^2 m_{0}^2)}\begin{aligned}[t]\biggl[&(D + F) (3 (a_{D} + a_{F}) (D + F) + 8 (\Delta a_{D} + \Delta a_{F})) M_{\pi}^2 \\
&-3 (a_{D} + a_{F}) (4 + (D + F)^2) m_{0}^2\biggr]\end{aligned} \\ 
&\quad+ \frac{I_{M}(M_{K})}{(36 F_{0}^2 m_{0}^2)}\begin{aligned}[t]\biggl[&2 (-\Delta a_{D} (D - 3 F) - 9 a_{F} (D - F)^2 + 3 a_{D} (D - F) (D + 3 F) + 3 (D + F) \Delta a_{F}) M_{K}^2  \\ 
&-3 (3 a_{F} - 6 a_{F} (D - F)^2 + a_{D} (3 + 2 (D - F) (D + 3 F))) m_{0}^2\biggr]\end{aligned} \\ 
&\quad+ \frac{I_{M}(M_{\eta})}{(24 F_{0}^2 m_{0}^2)}(a_{D} + a_{F}) (D - 3 F)^2 (m_{0}^{2} - M_{\eta}^{2}) \\ 
&\quad- \frac{I_{MB}(M_{\pi})}{(48 F_{0}^2 m_{0}^2)}(D + F) M_{\pi}^2 \left[-16 (\Delta a_{D} + \Delta a_{F}) (4 m_{0}^2 - M_{\pi}^2) + 3 (a_{D} + a_{F}) (D + F) (-8 m_{0}^2 + 5 M_{\pi}^2)\right] \\ 
&\quad- \frac{I_{MB}(M_{K})}{(36 F_{0}^2 m_{0}^2)}M_{K}^2 \begin{aligned}[t]\biggl[&2 (D (\Delta a_{D} - 3 \Delta a_{F}) - 3 F (\Delta a_{D} + \Delta a_{F})) (4 m_{0}^2 - M_{K}^2) \\ 
&+ 3 (D - F) ((a_{D} - 3 a_{F}) D + 3 (a_{D} + a_{F}) F) (-8 m_{0}^2 + 5 M_{K}^2)\biggr]\end{aligned}\\ 
&\quad+ \frac{I_{MB}(M_{\eta})}{(48 F_{0}^2 m_{0}^2)}(a_{D} + a_{F}) (D - 3 F)^2 M_{\eta}^2 (-8 m_{0}^2 + 5 M_{\eta}^2) \\ 
&\quad+ \frac{I_{MBB}(0,M_{\pi})}{(16 F_{0}^2 m_{0}^2)}(a_{D} + a_{F}) (D + F)^2 M_{\pi}^2 (8 m_{0}^4 - 12 m_{0}^2 M_{\pi}^2 + 3 M_{\pi}^4) \\ 
&\quad+ \frac{I_{MBB}(0,M_{K})}{(12 F_{0}^2 m_{0}^2)}(D - F) ((a_{D} - 3 a_{F}) D + 3 (a_{D} + a_{F}) F) M_{K}^2 (8 m_{0}^4 - 12 m_{0}^2 M_{K}^2 + 3 M_{K}^4) \\ 
&\quad- \frac{I_{MBB}(0,M_{\eta})}{(48 F_{0}^2 m_{0}^2)}(a_{D} + a_{F}) (D - 3 F)^2 M_{\eta}^2 (8 m_{0}^4 - 12 m_{0}^2 M_{\eta}^2 + 3 M_{\eta}^4) \\ 
&\quad- \frac{(D + F) M_{\pi}^4}{(2304 \pi^{2}F_{0}^2 m_{0}^4 )} \biggl[9 (a_{D}+a_{F}) (D + F) (2 m_{0}^2 - M_{\pi}^2)  + 4 (\Delta a_{D} + \Delta a_{F}) (6 m_{0}^2 - M_{\pi}^2)\biggr] \\ 
&\quad+ \frac{M_{K}^4}{(3456 \pi^{2}F_{0}^2 m_{0}^4 )} \begin{aligned}[t]\biggl[&54 a_{F} (D - F)^2 (2 m_{0}^2 - M_{K}^2) - 18 a_{D} (D^2 + 2 D F - 3 F^2) (2 m_{0}^2 - M_{K}^2) \\ 
&+ (D (\Delta a_{D} - 3 \Delta a_{F}) - 3 F (\Delta a_{D} + \Delta a_{F})) (6 m_{0}^2 - M_{K}^2)\biggr]\end{aligned}\\ 
&\quad+ \frac{(a_{D} + a_{F}) (D - 3 F)^2 (2 m_{0}^2 M_{\eta}^4 - M_{\eta}^6)}{(768 \pi^{2}F_{0}^2 m_{0}^4 )}\,, 
\end{split}
\end{align}
\begin{align}
\begin{split}\label{eq:last}
A^{v,8}_{pp} &= -Z_{N}\frac{a_{D}-3a_{F}}{2\sqrt{3}}-\frac{4}{3\sqrt{3}} \biggl[ 3 (t_{1}-3t_{2}) (2 M_{K}^2 + M_{\pi}^2)+2 (10 t_{3} - 6 t_{4} + t_{7} - 3 t_{8} + 6 t_{9}) (M_{K}^2 - M_{\pi}^2)\biggr]\\
 &\quad+ \frac{I_{M}(M_{\pi})}{(8F_{0}^{2}m_{0}^{2})}\sqrt{3} (a_{D} - 3 a_{F}) (D + F)^2 (-m_{0}^2 + M_{\pi}^2)\\
 &\quad+ \frac{I_{M}(M_{K})}{(12 \sqrt{3} F_{0}^2 m_{0}^2)}\begin{aligned}[t]\biggl[&2 (5 D \Delta a_{D} - 4 a_{D} D (D - 3 F) - 3 \Delta a_{D} F - 3 D \Delta a_{F} + 9 F \Delta a_{F}) M_{K}^2\\ 
 &+ (-27 a_{F} + a_{D} (9 + 8 D (D - 3 F))) m_{0}^2\biggr]\end{aligned}\\
 &\quad- \frac{I_{M}(M_{\eta})}{(24 \sqrt{3} F_{0}^2 m_{0}^2)}(a_{D} - 3 a_{F}) (D - 3 F)^2 (m_{0}^{2} - M_{\eta}^{2})\\
 &\quad+ \frac{I_{MB}(M_{\pi})}{(16 F_{0}^2 m_{0}^2)}\sqrt{3} (a_{D} - 3 a_{F}) (D + F)^2 M_{\pi}^2 (8 m_{0}^2 - 5 M_{\pi}^2)\\
 &\quad+ \frac{I_{MB}(M_{K})}{(6 \sqrt{3} F_{0}^2 m_{0}^2)}M_{K}^2 \begin{aligned}[t]\biggl[&(5 D \Delta a_{D} - 3 \Delta a_{D} F - 3 D \Delta a_{F} + 9 F \Delta a_{F}) (4 m_{0}^2 - M_{K}^2)\\ 
 &+ 2 a_{D} D (D - 3 F) (-8 m_{0}^2 + 5 M_{K}^2)\biggr]\end{aligned}\\
 &\quad+ \frac{I_{MB}(M_{\eta})}{(48 \sqrt{3} F_{0}^2 m_{0}^2)}(a_{D} - 3 a_{F}) (D - 3 F)^2 M_{\eta}^2 (8 m_{0}^2 - 5 M_{\eta}^2)\\
 &\quad+ \frac{I_{MBB}(0,M_{\pi})}{(16 F_{0}^2 m_{0}^2)}\sqrt{3} (a_{D} - 3 a_{F}) (D + F)^2 M_{\pi}^2 (8 m_{0}^4 - 12 m_{0}^2 M_{\pi}^2 + 3 M_{\pi}^4)\\
 &\quad- \frac{I_{MBB}(0,M_{K})}{(3 \sqrt{3} F_{0}^2 m_{0}^2)}a_{D} D (D - 3 F) M_{K}^2 (8 m_{0}^4 - 12 m_{0}^2 M_{K}^2 + 3 M_{K}^4)\\
 &\quad+ \frac{I_{MBB}(0,M_{\eta})}{(48 \sqrt{3} F_{0}^2 m_{0}^2)}(a_{D} - 3 a_{F}) (D - 3 F)^2 M_{\eta}^2 (8 m_{0}^4 - 12 m_{0}^2 M_{\eta}^2 + 3 M_{\eta}^4)\\
 &\quad- \frac{\sqrt{3} (a_{D} - 3 a_{F}) (D + F)^2 (2 m_{0}^2 M_{\pi}^4 - M_{\pi}^6)}{(256 \pi^{2}F_{0}^2 m_{0}^4)}\\
 &\quad+ \frac{M_{K}^4}{(1152 \sqrt{3} \pi^{2}F_{0}^2 m_{0}^4)} \begin{aligned}[t]\biggl[&- (5 D \Delta a_{D} - 3 \Delta a_{D} F - 3 D \Delta a_{F} + 9 F \Delta a_{F}) (6 m_{0}^2 - M_{K}^2) \\ 
 &+ 24 a_{D} D (D - 3 F) (2 m_{0}^2 - M_{K}^2)\biggr]\end{aligned}\\
 &\quad- \frac{(a_{D} - 3 a_{F}) (D - 3 F)^2 (2 m_{0}^2 M_{\eta}^4 - M_{\eta}^6)}{(768 \sqrt{3} \pi^{2}F_{0}^2 m_{0}^4)}\,. 
\end{split}
\end{align}
\end{widetext}
\section{Discussion and conclusion}
\begin{figure}
\subfigure[]{\includegraphics[width=0.45\textwidth]{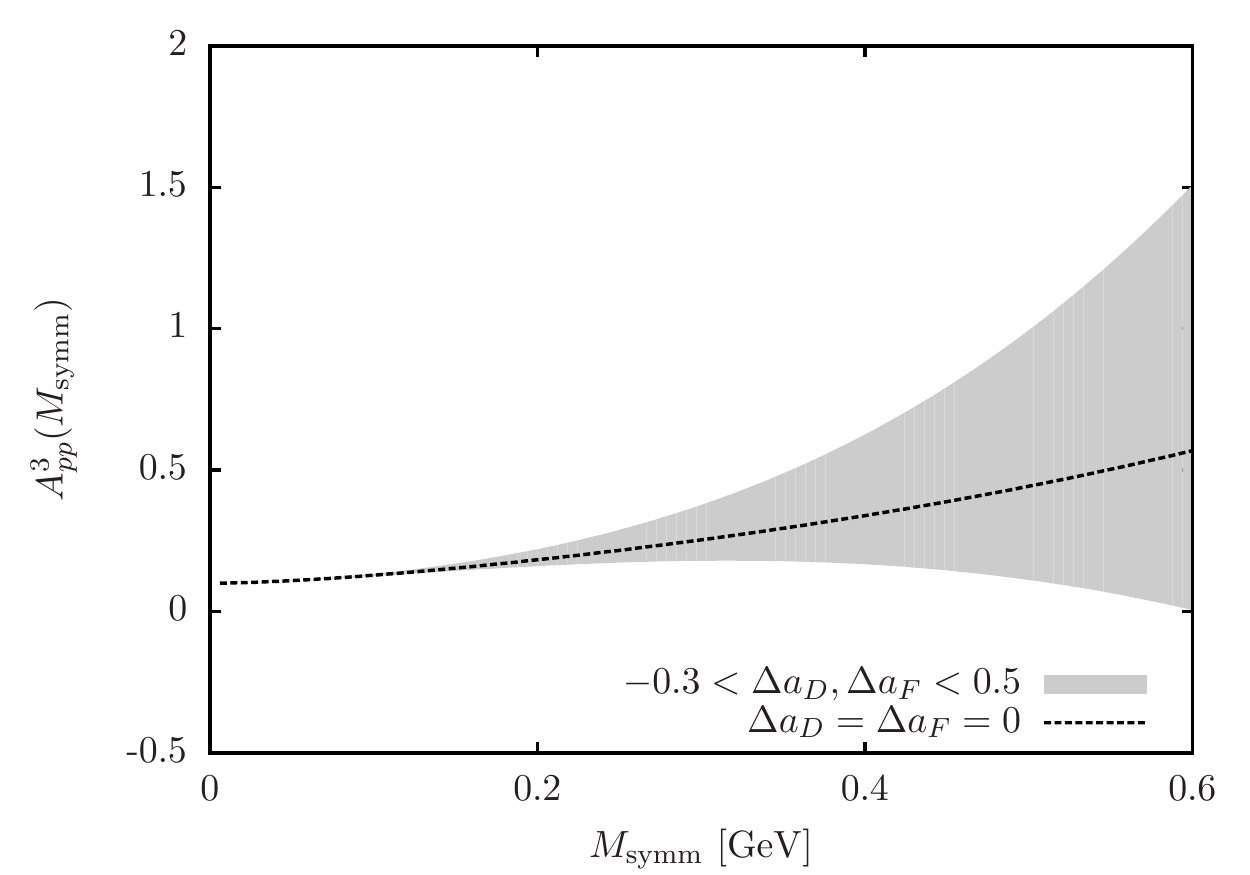}}\\\subfigure[]{\includegraphics[width=0.45\textwidth]{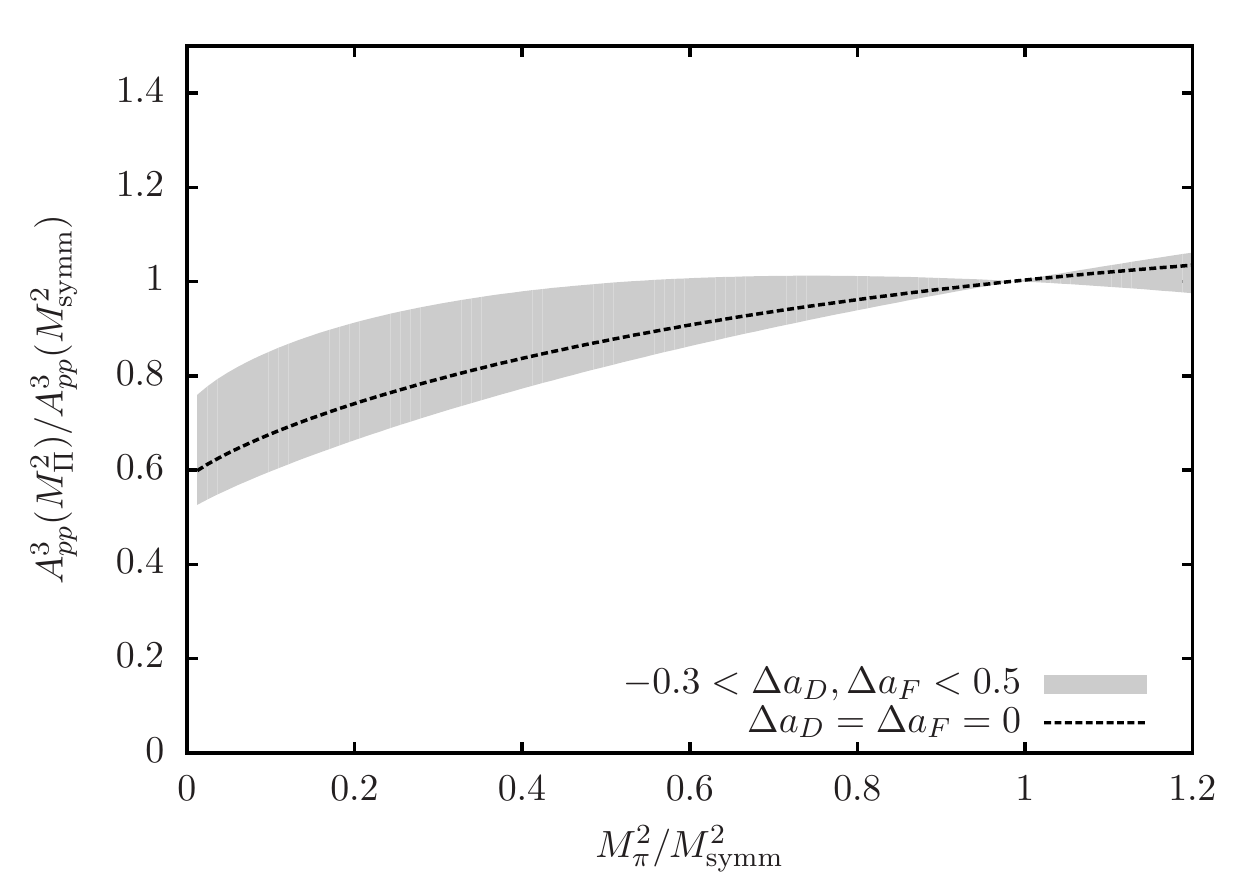}}
\caption{The form factors $A^3_{pp}(M_{\text{symm}})$ and $A^3_{pp}(M_{\pi}^2)/A^3_{pp}(M_{\text{symm}}^2)$ for different values for $\Delta a_D,\Delta a_F$. The expected size of higher-order corrections is represented by the grey band, generated as specified in the text, while the dotted line marks the result for  $\Delta a_D=\Delta a_F=0$.}
\label{fig:a3ppkonv}
\end{figure}

In this work, we have calculated the flavor-singlet and flavor-octet generalized form factor $A^{s,v}_{B'B}$ in the forward limit for three dynamical quark flavors. We have presented the results for the nucleon-channels  including the full loop functions, employing the covariant regularization scheme of \cite{Becher:1999he}. We have also calculated the matching relations between the $SU(3)$ parameters and the $SU(2)$ parameters introduced in \cite{Dorati:2007bk}. Together with the information from two-flavor lattice simulations and the corresponding results of \cite{Dorati:2007bk}, these relations can be used as constraints on fitting parameters when chiral extrapolations of three-flavor lattice calculations of the matrix elements in eqs. \eqref{eq:Ms}, \eqref{eq:Mv} are performed. 
However, a word of caution is in order here. It is well-known that three-flavor baryon ChPT is only poorly converging, mostly due to the large kaon and eta masses, see \textit{e.g.} \cite{Borasoy:1996bx,Mai:2009ce}. It is therefore reasonable to expect that a leading-one-loop calculation as presented here might not be sufficient for a satisfactory description of lattice data near the physical regime. \\
Despite this known difficulty, there exist some approaches to lattice simulations where the chiral extrapolation formulae derived here should be useful. The strategy followed in \cite{Bietenholz:2010jr, Bietenholz:2010si} is to start from the $SU(3)$-symmetric point where $m_{u}=m_{d}=m_{s}$ and to approach the physical point keeping the average (singlet) quark mass fixed. For realistic values of the quark masses, this corresponds to octet meson masses of $M_{symm}^{2}=\frac{1}{3}(2M_{K}^{2}+M_{\pi}^{2})\approx (411\,\mathrm{MeV})^{2}$. So, in the vicinity of the symmetric point, the kaon and eta masses are considerably smaller than in the real world. As one can easily see from eq. \eqref{eq:Lag2}, only the constants $t_1,t_2,t_{11}$  are relevant at the symmetric point, while the other couplings only parametrize deviations from the $SU(3)$ symmetry limit. In the case of baryon masses, it has been argued in \cite{Bietenholz:2011qq} that an accurate extrapolation to the physical point can be obtained already employing an expansion linear in the symmetry breaking due to the quark masses, which amounts to a linear extrapolation in $M_{symm}^{2}-M_{\pi}^{2}$. This seems to work reasonably well and leads to so-called fan plots for the baryon masses, shown in \cite{Bietenholz:2011qq}. In the case of GPDs, however, one should expect that it is more important to take the proper chirally nonanalytic behaviour for small pion masses into account, because already the leading  correction of $\mathcal{O}(p^2)$ to these form factors features chiral logarithms. This is not the case for baryon masses, where the leading quark mass correction is just given by mass insertions from local counterterms in the chiral Lagrangian.\\
In the following, we would like to argue that the mentioned type of lattice simulations is well suited for an application of our extrapolation formulae, presented in the last section. To get a rough estimate of the size of possible higher-order corrections to these results for this case, we consider the contributions from the graphs in figs. \ref{fig:ga} and \ref{fig:gb}, which turn out to start only at next-to-leading loop order $\mathcal{O}(p^3)$. They are proportional to the couplings $\Delta a_{D,F}$, of which only the order of magnitude is (roughly) known, see table 2 of \cite{Dorati:2007bk} for the value of the corresponding $SU(2)$ parameter $\Delta a^v_{2,0}$. We vary these couplings in the range $-0.3<\Delta a_D,\Delta a_F<0.5$, with all other couplings fixed at typical values consistent with the analysis in the $SU(2)$ sector and earlier work on BChPT ($a_D=a_F=0.1$, $m_0=1\,\text{GeV}$, $D=0.8$, $F=0.5$ and $F_{0}=0.09\,\text{GeV}$) and all $t_i$ set to zero. The generated band shown in fig. \ref{fig:a3ppkonv} should give a good impression of the expected size of higher-order contributions.
In the lower panel of fig. \ref{fig:a3ppkonv}, the singlet quark mass is kept fixed, while the pion mass is varied, in analogy to the ``fan plots'' shown in \cite{Bietenholz:2011qq}.
Regarding the rather generous range for the variation of the LECs $\Delta a_D,\Delta a_F$, and earlier experiences with the convergence properties of three-flavor chiral expansions, the band in the lower panel indicates an improved stability to higher order corrections, which leaves us optimistic that the presently derived chiral extrapolation formulae provide a realistic description for the special type of lattice simulations described above. 
In principle it is straightforward to extend the analysis to a next-to-leading one-loop calculation. On the other hand, some additional coupling constants will appear if one pushes the calculation to higher orders, and one would have to specify the individual baryon masses in the loop functions, which complicates the calculations a lot. Even at the order we are presently working, it will require an enormous amount of three-flavor lattice data to be able to pin down all of our LECs $t_{i}^{(r)}$. Therefore, we defer the extension of the present results to a later communication. In our opinion, the most promising strategy for the moment is to stay with the results as derived here, and to first concentrate on a better understanding of the two-flavor sector, e.g. by analyzing higher-order corrections to the existing calculations \cite{Dorati:2007bk,Chen:2001et,Arndt:2001ye}. There, the number of new undetermined parameters is much more limited, and convergence is a less severe issue than in the three-flavor framework. Work along this direction is already in progress \cite{Dorati:2010bk,wir:2022xx}.  
\begin{appendix}
\begin{widetext}
\section{Loop functions}
In sec.~\ref{sec:res}, we have used the following abbreviations for the loop functions:
\begin{align}
I_{M}(M) = \frac{M^{2}}{8\pi^2}\ln\alpha\,,
\end{align}
where $\alpha = M/m_{0}\,$. Note that we use $\mu=m_{0}$ for the renormalization scale everywhere, following the original proposal of Infrared Regularization by Becher and Leutwyler. Furthermore,
\begin{align}
I_{MB}(M) = \frac{1}{16\pi^{2}}\left[(2\ln\alpha - 1)\frac{\alpha^{2}}{2} + \alpha\sqrt{4-\alpha^{2}}\arccos\left(-\frac{\alpha}{2}\right)\right]\,.
\end{align}
Moreover, the renormalized three-point function in Infrared Regularization, taken at $\Delta^{2}=0$, is given by
\begin{align}
I_{MBB}(0,M) = -\frac{1}{32\pi^2m_{0}^{2}}\left[2\ln\alpha + 1 -\frac{2\alpha}{\sqrt{4-\alpha^{2}}}\arccos\left(-\frac{\alpha}{2}\right)\right]\,.
\end{align}
The nucleon wave function renormalization factor, at the one-loop level, is given by
\begin{align}
\begin{split}
Z_{N} &= 1-M_{\pi}^{2}\frac{3 (D + F)^2}{(32 \pi^2 F_{0}^2)}-M_{K}^{2}\frac{5 D^2 - 6 D F + 9 F^2}{(48 \pi^2 F_{0}^2)}-M_{\eta}^{2}\frac{(D - 3 F)^2}{(96 \pi^2 F_{0}^2)} \\
 &- \frac{3 (D + F)^2 M_{\pi}^3 (-3 m_{0}^2 + M_{\pi}^2)}{(16 \pi^2F_{0}^2 m_{0}^3 \sqrt{4 - \frac{M_{\pi}^2}{m_{0}^2}})}\arccos\left(-\frac{M_{\pi}}{2m_{0}}\right)- \frac{(D - 3 F)^2 M_{\eta}^3 (-3 m_{0}^2 + M_{\eta}^2)}{(48 \pi^2F_{0}^2 m_{0}^3 \sqrt{4 - \frac{M_{\eta}^2}{m_{0}^2}})}\arccos\left(-\frac{M_{\eta}}{2m_{0}}\right)  \\
 &- \frac{(5 D^2 - 6 D F + 9 F^2) M_{K}^3 (-3 m_{0}^2 + M_{K}^2)}{(24 \pi^2F_{0}^2 m_{0}^3 \sqrt{4 - \frac{M_{K}^2}{m_{0}^2}})}\arccos\left(-\frac{M_{K}}{2m_{0}}\right)+ \frac{3 (D + F)^2 M_{\pi}^2 (-3 m_{0}^2 + 2 M_{\pi}^2)}{(32 \pi^2 F_{0}^2 m_{0}^2)}\log\left(\frac{M_{\pi}}{m_{0}}\right)\\
 & + \frac{(5 D^2 - 6 D F + 9 F^2) M_{K}^2 (-3 m_{0}^2 + 2 M_{K}^2)}{(48 \pi^2 F_{0}^2 m_{0}^2)}\log\left(\frac{M_{K}}{m_{0}}\right)+ \frac{(D - 3 F)^2 M_{\eta}^2 (-3 m_{0}^2 + 2 M_{\eta}^2)}{(96 \pi^2 F_{0}^2 m_{0}^2)}\log\left(\frac{M_{\eta}}{m_{0}}\right)\,.
\end{split}
\end{align}
\end{widetext}
\end{appendix}
\acknowledgments{This work was supported by SFB/TR-55.}
\bibliographystyle{apsrev}
\bibliography{bibliography}
\end{document}